# SCALABLE GRID RESOURCE DISCOVERY THROUGH DISTRIBUTED SEARCH

Fouad Butt[1], Syed Saadat Bokhari[2], Abdolrezza Abhari[3] and Alexander Ferworn[4]

Department of Computer Science, Ryerson University, Toronto, Ontario

[1]f2butt@ryerson.ca
[2]saadat.bokharisyed@ryerson.ca
[3]aabhari@scs.ryerson.ca
[4]aferworn@scs.ryerson.ca

## ABSTRACT

*This paper proposes a simple and scalable web-based model for grid resource discovery for the Internet. The resource discovery model contains the metadata and resource finder web services. The information of resource finder web services is kept in the repositories that are distributed in the application layer of Internet. The resource finder web services will be discovered by sending queries to the repositories in a similar way as the DNS protocol. The underlying technology for implementation of the two architectures of this model is introduced.*

*These architectures are: Direct and Centralized Web-Based Grid Resource Discovery. The resource discovery time is computed after simulating each of these models in GridSim.*

*By performing scalability tests, we found that when increasing the load on the grid with more users and resources, the cost of our model in comparison to the grid resource discovery time is marginal.*

## 1 INTRODUCTION

With the popularity of the Internet, there is a strong need for grid computing applications to share their resources through the Internet. Therefore, it is important for a grid resource discovery model to feature certain scalability characteristics for finding and accessing the resources. However, most of these grid resource discovery models do not focus on scalability when the goal is accessing the grid resources through the Internet. Considering the importance of scalability for accessing grid resources through the Internet, we address two separate problems:

1. Providing a scalable solution: A scalable solution should be provided with respect to all three issues mentioned by Nueman [1]. The scalable solution is required for providing Internet access to grid resources. This solution should remain effective when the number of users and resources increase along with geographical expansion of the system. This solution should also be capable of addressing the administrative information requirements when the system is growing. Specifically, this scalable information administration means providing effective methods for searching and storage of the information of the grid with increasing resources in Internet.

2. Examining the scalability of the solution: Another problem in our work is that we need to test the scalability of our model by varying the number of users and resources and examining the effectiveness of information administration of the model. To accomplish this task, proper performance metrics must be utilized.

To address the first problem, we believe that using web services can solve the geographical expansion problem and provide a scalable and reliable interface for Internet users to access grid resources. Web services are not dependent on any computing platform or programming language and instead use Simple Object Access Protocal (SOAP) [2] messages with Extensible Markup Language (XML) format [3] for communication as described by Coulouris et al., in [4]. Our solution is an architectural model that employs Web services as the user interface to provide scalable Internet access to grid resources.

To address the second problem, we found that testing the scalability for the real grid on the Internet is a hard and expensive task. Thus, we need to simulate the grid to be able to examine its scalability. Grid simulation is also a difficult task because a grid is not in the form of a typical network. Network simulation tools such as ns2 [5] are not useful for grid simulation; therefore, we need to find suitable grid simulation tools.

The model is similar to Domain Name Service (DNS) protocol and uses the same techniques such as caching and distributed search for providing scalability. Our model uses queries for searching the Uniform Resource Locator (URL) of the web services in a similar manner as DNS. Instead of name servers, our model uses repositories with similar zone files for keeping the information of the URLs of the resource finder web services and their related grid resources.

We implement the model in GridSim. Only the best case scenarios of our model are tested for regional searches because of the complexity of DNS-based distributed search in the model.

Our architectural model uses caching techniques in its repositories similar to the DNS protocol. It first searches for resources from the regional repository, and if the required resource is not found, then another repository outside of the regional domain will be contacted.

According to the assumptions, we do not discuss the structure of the resource finder web services and we do not discuss the internal structure of Universal Description, Discovery and Integration (UDDI) in this work. We also assume a tree structure of repositories with a hierarchy similar to the domain name services in DNS protocol is available. We will introduce the underlying technology to implement our model for the resource discovery scenarios but we will not go into the details of the whole hierarchical model of repositories. The rest of the paper is structured as follows:

Section 2 outlines other works related to resource discovery. Section 3 is a discussion of the implemented model, while section 4 provides detailed results supported by the statistical analysis in section 5. Section 6 contains concluding remarks and directions for further research.

## 2 RESOURCE DISCOVERY BACKGROUND

The sharing and integration of heterogeneous resources for computational and data storage projects has become an important area of grid research. Resources on grid systems are widely distributed and heterogeneous in comparison to traditional and cluster systems. Resource discovery is a key grid management tool for extraction of resource information in a grid environment. Resource discovery management in a grid is based on resource organization. Most approaches to resource discovery in grid computing treat resources equally. These approaches face problems in terms of efficient response times and the need to generate complex queries. Therefore, a simple yet scalable resource discovery architecture is required. Previous research pertaining to the subject of resource discovery is discussed next.

### 2.1 Resource Discovery in HealthGrids

Naseer and Stergioulas describe an approach utilizing web services to facilitate resource discovery in a particular type of grid network: a HealthGrid [6]. The authors define a HealthGrid as a specialized network for the purpose of provisioning and supporting healthcare services [6]. Their contribution focuses on the problem of resource discovery in HealthGrids and constitutes a web service-based resource discovery model [6]. The model is categorized into two levels: the operational level and management level [6]. The operational level consists of various core services that are especially relevant to healthcare processes, which include: scanning and data collection; modeling and simulation; visualization; and, comparison and analysis services [6]. The management level forms a collection of services that ensure registration, configuration, scheduling and discovery of operational level services [6].

Several problems arise within the context of a HealthGrid when a service-oriented architecture is implemented, which require solutions from a healthcare perspective [6]. For instance, to address the heterogeneous nature of medical data and environments, the authors suggest definition of biomedical standards that encompass the various data types and environments to present a unified view of heterogeneous resources [6].

Simple Object Access Protocol (SOAP) messages support communication between services, which can be written in extensible markup language (XML) or rich data format (RDF) [6]. The authors also note the applicability of Web Service Definition Language (WSDL) for web service definition [6].

## 2.2 Ant Colony-Based Distributed Resource Discovery

Recent work by Brocco, Malatras and Hirsbrunner follows a distributed approach to the problem of resource discovery in a self-structured grid [7]. The authors employ an ant colony algorithm to form and support a peer-to-peer (P2P) overlay network of all the grid nodes with a minimal number of links and a restricted network size [7]. Six types of ants participate in the ant colony overlay formation and optimization process: discovery, construction-link, optimization-link, unlink, update neighbours and ping ants [7].

Discovery ants perform random walks across the network for a limited amount of time and store collected information about the visited nodes in the alpha-table of their originating node [7]. Construction-link ants originate from new peers that want to enter the overlay network [7]. If a destination node is at maximum capacity, the ant is forwarded to survey its neighbouring nodes and once a suitable node is found, it is added to the neighbourhood set for the originating node [7]. The destination node is also added to the neighbourhood set of the originating node [7]. Optimization-link ants also form connections between nodes but only if the connection meets the optimization requirements defined by a connection rule [7]. Additionally, optimization-link ants cannot be forwarded to neighbouring nodes [7]. Unlink ants remove any existing links between two nodes if a disconnection rule applies or, if a node leaves the network [7].

To increase efficiency, they also proactively situate nodes with similar capabilities as the requesting node and cache them to reduce network overhead [7]. The proactive caching is accomplished in a similar as a gossiping algorithm by periodically gathering information about resources similar to the requesting node and storing it in a local cache [7].

Their current algorithm, BlatAnt-S, is compared with their previous algorithm, which is BlatAnt-R [7]. Drawing from this comparison the authors note marked improvements in hit rate [7]. Furthermore, with respect to the overlay network, average path length is constrained and the degree of each node is minimized along with the number of unnecessary links [7]. Network bandwidth requirements are also reduced due to caching [7].

## 2.3 Ontology-Based Resource Discovery

To improve resource discovery in the Open Grid Services Architecture (OGSA), Pahlevi and Kojima augment the framework suggesting an ontology-based approach [8]. The authors claim that due to a lack of semantic constraints on the information pertaining to service instantiations, sharing this information becomes difficult [8]. Furthermore, it is also complicated to process information by utilizing automated tools [8]. In support of their claims, the researchers add semantics to the service information by defining certain domain-specific ontologies [8]. Their results point to improvements in search results and lend to the possibility of automated resource discovery.

## 2.4 Layered Architecture-Based Resource Discovery

The layered architecture of grid computing was proposed by Foster et al. [9] in 2001. It includes five layers: Fabric, Connectivity, Resource, Collective and Application layers. The Fabric layer is the first layer, handling local resources at a particular site. The Connectivity layer uses network communication protocols for data transfer between resources. The Resource layer uses protocols of the connectivity layer and interfaces provided by the fabric layer to manage a single resource. The collective layer holds the information of multiple resources and manages resource discovery, task scheduling and the allocation of services. The collective layer discovers resources based on the information from lower layers. Finally, the application layer is the closest layer to the user and provides applications within the virtual organization access to the grid.

## 2.5 A Hierarchical Model for Resource Discovery

Yin et al. [10] propose a model called the Hierarchical Resource Organizational Model. The model consists of three layers to process the information in a grid. These three layers include: Physical Network, Resource Information and Index Information. The Physical Network Layer is at the lowest level containing the physical resources linked with each other on the Internet. For each resource, a resource node is placed in the Resource Information Layer, which lies in the middle. Therefore, the middle layer contains virtual organizations (VO), which is a group of resource nodes in a star topology with a super node in the center. The super node stores all the information regarding the resources of a VO as adjacent lists. Forming a ring topology, super nodes lie in the Index Information Layer, which is the highest level for resource discovery. This layer stores information pertaining to all super nodes of the middle layer and composition of the multiple layers form the basis for hierarchical resource discovery as presented in [10].

Varying the number of resource nodes, Yin et al. [10] perform three simulation tests on three types of discovery models. They compare their model with two other models that they called Exhaustive and Lumped models. The Exhaustive model searches all the resource information nodes and the Lumped model uses a single point for resource discovery.

As explained in [10], the hierarchical model not only outperforms the other two models in terms of resource discovery time but it also does not exhibit performance problems of the other two models. However, single point-based resource discovery in the Lumped Model can become a bottleneck and source of failure. Also, blind search discovery--as in the Exhaustive Model--does not scale when the resources are increasing.

## 2.6 Tree Structure-Based Resource Discovery

Sun et al. [11] propose Resource Category Tree (RCT) that organizes the resources as an AVL tree. AVL trees are self-balanced structures in that their height is adjusted when nodes are added dynamically to the tree as described by Horowitz et al. in [12]. Applying this method, the authors argue that flexibility and scalability for grid resource discovery is improved. To satisfy resource queries, resource characteristics are assigned to each resource. The important characteristics that represent the resources are called primary attributes (PA) in [11]. These attributes are organized as tree nodes, to handle range-based queries. A query that includes multiple attributes within a range is called a multi-attribute range query.

The RCT proposed by Sun et al. [11] is simulated and compared with a hierarchical structure. The simulated RCT and hierarchical model each consist of 100 nodes. The query load per node and average search length (ASL) are performance metrics for this simulation. ASL is the average number of nodes that a request traverses to process a query. The RCT is found to be 50% more efficient in terms of the number of queries reaching each node than the hierarchical model. Regarding comparison of average search length, the results show that RCT searches between 8.5 to 9 nodes, while the hierarchical scheme searches all 100 nodes. Thus, it is concluded that RCT is an efficient and complete solution for resource discovery [11].

## 2.7 Grid Resource Discovery Based on Semantic Information

A Peer-to-Peer (P2P) network consists of computers that share their resources with each other without using a server computer. The P2P grid model is similar to a P2P network that consists of a set of super nodes also called grid peers, as described by Xiong et al. in [13]. A grid peer manages a set of nodes. A node is a computer managing a group of local grid resources. When a user searches for resources, the grid peer domain of local resources is queried first. If the query is unsuccessful, the local grid peer redirects it by performing a random walk to the closest grid peers. The P2P system consists of autonomous agents that are able to accomplish unsupervised actions. These include the Request Agent and the Broker Agent. The Request Agent is responsible for handling queries and collecting results from a grid peer. If it cannot find the information, it sends a request for information to the Broker Agent of the same grid peer. The Broker Agent contains a database of resource information to be discovered from other grid peers, called the Global Knowledge Database.

As mentioned by Beheshti and Moshkenani [14], a grid computing architecture is required to discover the resources, since a large portion of the time in grid computing is spent on resource discovery. This time can be minimized through the addition of architectural layers. These layers manage all the resources of the grid.

The Galaxy architecture model proposed by Beheshti and Moshkenani [14] is a hypothetical grid architectural model backed by the idea of a universal grid. It consists of four levels of grid computing in which different forms of indexing called metadata and meta- metadata are employed for resource discovery.

These metadata, together with agent programs, are distributed in all layers of the grid. The metadata in Galaxy contains semantic information about the availability of resources that is updated by smart agent programs. In Galaxy, it is assumed that service-oriented programs are available at all levels of the grid and are called by agents when a resource change event is triggered. The metadata contains the latest resource information in XML format. At the core of Galaxy is the agent and software services, which update semantic information and assist with resource discovery.

The semantic information includes its ontology (i.e., the meaning and interpretation of information) that is needed for communication between different services. In Galaxy, the administrative information of the resources of any node is created and updated by agent programs and kept in metadata. The Galaxy layer for resource management is on top of the data link layer and under the network layer, which adds additional information into IP packets.

The use of web services with semantic information, multi-agent systems and metadata as declared by Beheshti and Moshkenani [14] will provide up-to-date resource information to the end user. Although no simulation work is done, it is predicted that resource access with the Galaxy architecture will be much faster.

## 2.8 UDDI-Based Resource Discovery

As described by Foster et al. [15], the Open Grid Service Architecture (OGSA) is the standard for integration of web services and grid computing. It introduces the concept of grid services and all the mechanisms required for using grid services such as naming, grid service description and grid service discovery. OGSA is a result of the Globus toolkit that was introduced by Foster et al. in [16]. The Globus toolkit is developed to use the grid services in practice. OGSA is evolving middleware that uses specifications of Web Service Resource Framework to build a web service-based grid.

Benson et al. [17] focus on internal structure of UDDI for discovering grid services. UDDI is used as the resource discovery mechanism of OGSA-based grids. However using UDDI for grid service has some problems because it is designed to be used for business services. According to Benson et al. [17], a UDDI registry keeps track of a resource through a string reference key. The UDDI has three basic design issues, which hinder its application towards grid services:

1. Missing of explicit data type for UDDI directory

2. Difficulties in handling regularly updating dynamic information such as continuous numeric type of CPU load which changes at instances

3. Limited query capability

Benson et al. [17] proposed a new UDDI centralized model of grid resource discovery with following modifications:

1. The issue of explicit data type, which does not exist in UDDI registry, is resolved by proposing the continuous variables of numeric type in UDDI registry.

2. The issue of dynamic information is resolved by introducing a new variable called lastUpdateTime in the UDDI registry for storing periodic update from resource providers.

3. The issue of limited query model is resolved by associating performance data like CPU load or machine attributes with a reference key.

Therefore, UDDI can be used for grid service discovery but with the above modifications. In this work, experiments are done to find the performance of a modified jUDDI under the system load measured by the update frequency of grid resources. It is mentioned by Benson et al. [17] that the implementation of UDDI version 3 will match the requirements for grid services.

## 2.9 Comparison with Other Architectures for Grid Resource Discovery

Our work is based upon web services and the simulation package GridSim that simulates a layered architecture for grid resource discovery. GridSim simulates resources, users, jobs and other grid components. Regarding the use of a central approach for discovering web services for grid computing, our work is similar to that presented by Benson et al. [17]. We use UDDI but we do not focus on the internal structure of UDDIs for discovering grid services.

However, we assume a required UDDI technology exists according to that proposed by Benson et al. [17]. Therefore, we are able to use UDDIs for discovering resource finder web service. We also assume the resource finder web service in our work has the requirements of a grid service that are suggested by Foster et al. [15] for OGSA but we do not focus on its internal structure.

Our model is a simplification of the upper two layers of the layered architecture model discussed in [9]. The application layer of our model provides a web interface for the user and the collective layer of our model is a web service focusing on resource discovery. They are therefore referred to as Web Interface and Resource Finder Web Service layers respectively and shown in Figure 1.

We do not address the rest of the three lowest layers of the layered architectural model presented by Foster et al. [9] but a combination of all of them is shown as the Grid Computing Resources/Nodes layer in Figure 1. Utilizing web services and XML-based metadata for grid computing is also proposed by Beheshti and Moshkenani [14] in the Galaxy architecture.

Considering the implementation of our model regarding TCP/IP, our model can be implemented in a way that is similar to the DNS protocol presented by Giordano [18]. The additional layer for our model could be placed in the application layer as Web Interface, and the SOAP messages would transfer between the resource finder and client web services. There can be many resource finder repositories in our model similar to Name Servers in DNS that can also be used for sending queries to other resource finder repositories to find requested resources.

We believe that by limiting the web services for communication between user and resource finder components and simplifying the metadata, our model can be implemented as part of the application layer. Therefore, we have solved the main drawback of Galaxy, which is impracticality.

## 3 ARCHITECTURAL MODEL AND IMPLEMENTATION STRATEGY

This section discusses our model. The discussion about the web-based architecture will be presented in subsection 3.1. This section also elaborates on the implementation strategy for our web-based layered architectures in subsection 3.2. Finally, conclusions are presented in subsection 3.3.

### 3.1 Web-Based Architecture for Grid Resource Discovery

Resource discovery is a time-consuming part of grid computing. To implement more effective resource discovery, we propose a model to employ web services in the upper layers of grid computing. Although our model may be considered to be a simplification of some of the previous models presented by Foster et al. [9] and Beheshti and Moshkenani [14], the structure and function of our model differ significantly from these models. This is explained in further detail in this section. The structure and function of our model is shown in Figure 1.

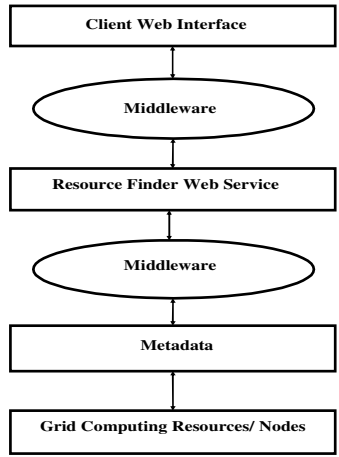

Figure 1. Architectural model

In our model, a user will access web services through the highest layer of the grid to locate the resources with the help of a resource finder (or resource broker) and metadata. The web services will then contact the resource finder to obtain the address of a resource. If the requested resource is available within the metadata, the resource finder will send the metadata related to that resource through the web service to the user. Metadata is the highest level of resource information and includes addresses in the grid.

To discover the resource finder service on the Internet, the client can send queries to repositories similar to Name Servers in DNS [19] as proposed by Giordano [18]. Metadata contains the highest level of information of the group of resources available within the local network or within the same region containing special criteria. The user who requests the resource may know the URL of the resource finder (e.g., by using a search engine or being in the same region) or may be redirected to send DNS queries to the repositories (UDDI) of the resource finder services to find the URL of the related resource finder.

Once the URL of the resource finder that has that resource on its list has been found, it can be accessed by the user through a related web service. Thus, in our model, end users will use web services in both of the above mentioned cases only to communicate with a resource finder that has the list of the group of the requested resources in its metadata.

We have submitted a paper discussing the original results of the implementation of the first version of the proposed model [20]. In this work, the model was modified extensively and shortcomings of the original work are addressed and the new results are included. The modified layers in our model are client web interface and resource finder web services.

The resource finder service searches metadata and returns a list of the available resources of the grid to the user by using common protocols within the application level of TCP/IP such as HTTP. Our model proposes XML format for metadata to be used by web services. As mentioned in section 2.9, our model could be considered a simplified form of application and collective layers of the layered architecture presented in [9]. Similar to the Galaxy model presented by Beheshti and Moshkenani [14], our model uses web services to find grid resources.

However, our model does not use any semantic components or smart agents because we focus on practicality and simplicity. The search for the resource finder and metadata components of our model can be considered similar to the distributed search model in Name Service (DNS) protocol presented by Mockapetris [19], with slight modifications that are explained next.

The simplest scenario is a regional resource finder that is known to the user in advance that has the resources requested by the user and acts as the broker to allocate these resources within the grid network. However, there are situations that are not that simple.

First, we consider a scenario in which the resource finder realizes there are many resources eligible to satisfy users search criteria. In that case a resource finder communicates with the user through a client web interface to ask the user to select the required resources from the list.

Second, we consider a scenario where the resource finder, by searching metadata, realizes there is no resource or a low number of resources available for the user. In that case, the user's search query will be sent to a regional registry service that contains the list of resource finders and high level information on their metadata. The search for a resource finder that has the resources required by the client continues in a distributed manner. The final result of this search in our model is the URL of the appropriate resource finder service that has the requested resources. The result of this search will be returned to the client. This approach is described by Giordano [18].

Since resource finders are web services, we assumed that similar to the approach proposed by Giordano [14]. DNS queries are used to discover the URL of the appropriate resource finder service by contacting central registries of each region in which web services are registered. This is a traditional method of using UDDI to find web services.

We used the distributed search method in the model not only because of the scalability of this approach, but also because this method is used in the model to update the information of resource finders registries based on the latest information in the grid. When the distributed search for resource finders is in progress, each resource finder registry in the path finally receives the information of the appropriate resource finder that has the requested resources.
Each of these resource finder registries in the path adds this new information to its list of registered services, similar to DNS name servers. Thus by employing this approach, not only do resource finder's metadata contain the latest information of the resources in their own region but the resource finder registries in the long run can also cache the recent information of other registered resource finders and their resources.

After finding the URL of the appropriate resource finder that has the list of required resources, the resource finder performs the duties of a resource broker by allocating the resources within the grid. It will submit the jobs to these resources, and after a job executes, it will deliver the results to the user.

### 3.2 Underlying Technology for Implementation of the Web-based Architecture

Our main goal throughout the implementation of the model is to develop a prototype that can be used together with grid simulator software to estimate any changes in grid resource discovery time as a result of adding our model to the grid. For implementation of the resource finder, we used a web service. The user access a client web interface implemented as a web application, which calls a resource finder service.

For simplicity, we assumed that the scenarios given next are within the regional search for a resource finder service. Due to the complexity of distributed search, it was not implemented in its entirety in our model. The implementation for two scenarios of regional search is as follows:

First, we implemented a model in which the resource finder service end point (URL) is known to the client and the resource finder has all the resources. Since in this approach there is no need to discover a resource finder service, and a client can call the resource finder service directly, this approach is called Direct Web-Based Grid Resource Discovery model (DWGRD). The implementation of DWGRD is elaborated upon in section 3.3.

In the second implementation approach, the resource finder web services are registered in a regional central registry, which is a part of our web application. In this approach, the user knows the address of a registry location such as UDDI, by which they can find the URL of the resource finder service that has the required resources. The discovery of a proper resource finder by using UDDI is added to the DWGRD model and the resulting architecture is called the Centralized Web-Based Grid Resource Discovery (CWGRD) model. Once the service is found, the client then sends the request of requirements to the service, and the result is returned to the client. CWGRD is discussed in section 3.4.

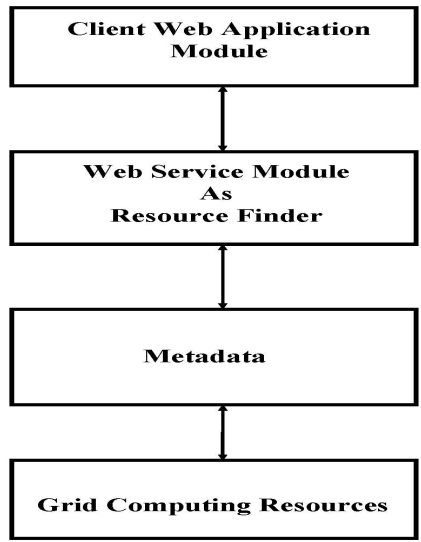

Figure 2. Implementation Flow of DWGRD

### 3.3 Direct Web Service-Based Grid Resource Discovery

For implementation of the DWGRD model, NetBeans IDE version 6.5.1 is used because of its strength in the implementation of web service applications [21]. The resource finder web service is accessed by a web application called Client Web Application (CWA). The resource finder web service calls GridSim to simulate accessing the resources in the grid environment.

The messages are then exchanged between the web service and CWA using SOAP. The web service and CWA are executed on the same server. The implementation flow of the DWGRD model in NetBeans IDE 6.5.1 is shown in Figure 2 and simulation results are given in section 4.

The implemented software has two parts:

1. The CWA has a Java Server Page (JSP) page as an interface that sends entered parameters to the web service application. The web service module acts as a grid resource finder. The CWA receives timing results from the grid resource finder after the next step.

2. The grid resource finder web service calls the Grid Information Service class of GridSim, which simulates grid entities such as users, resources and jobs in the grid environment.

### 3.4 Centralized Web Service-Based Grid Resource Discovery Implementation

The CWGRD model is also implemented as a web application in NetBeans IDE Version 6.5.1. The UDDI registry technology used is called jUDDI. The detailed explanation of installation of jUDDI in Apache Tomcat Server is given in [22].

We are using UDDI and jUDDI registries while assuming they have been enhanced by the solution provided in [17] and are capable of discovering grid resources. The jUUDI server is used for our model for the following reasons:

1. It keeps the information in XML format and it compatible with web service technology.

2. It keeps all the details required for discovering resource finder web services.

3. We use multiple UDDIs distributed in Internet instead on the central UDDI. It means we will use multiple jUDDI servers at many regional bases similar to DNS servers. These jUDDI servers will be communicating with each other using distributed search for resource discovery similar to DNS servers. Using multiple jUDDI servers, we will not have the problem of a central bottleneck when having only one repository.

Once the service is registered into the jUDDI database that is connected to the databases server, the service can be accessed from the registry using CWA. The CWGRD model is implemented as shown in Figure 3.

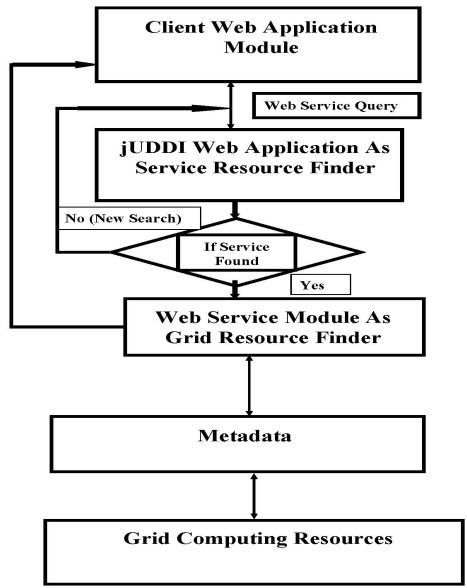

Figure 3. Implementation Flow of CWGRD

The CWA is constructed in such a way that it sends the query to find a service from the jUDDI registry. The result of the service query is returned and the time in milliseconds that it took to satisfy the query from the jUDDI registry is recorded. This time will be added to the time reported by DWGRD that is determined by the resource finder service.

## 4 SIMULATION RESULTS

GridSim simulates grid entities such as users and resources with their registration time. GridSim (No Architecture) is a simple java application that executes in NetBeans IDE 6.5.1 without any use of web services. It acts as a base for DWGRD and CWGRD models. The grid resource discovery time found in GridSim is less than both DWGRD and CWGRD. However, when the number of resources and users is increased, the discovery time of GridSim becomes closer to the discovery times of two models.

The main goal of performing these tests is to assess the scalability of the system. We want to see after which number of resources and users does the difference between the cost of each model and GridSim become negligible. Although there is a difference between the average resource discovery times of the three models, we can observe that this difference decreases under a higher load on the grid. In other words, the difference between mean resource discovery times utilizing no architecture (GridSim) and either of the models (CWGRD or DWGRD) decreases as strain on the grid is increased. This phenomenon can be observed in Figures 4.1 and 4.2.

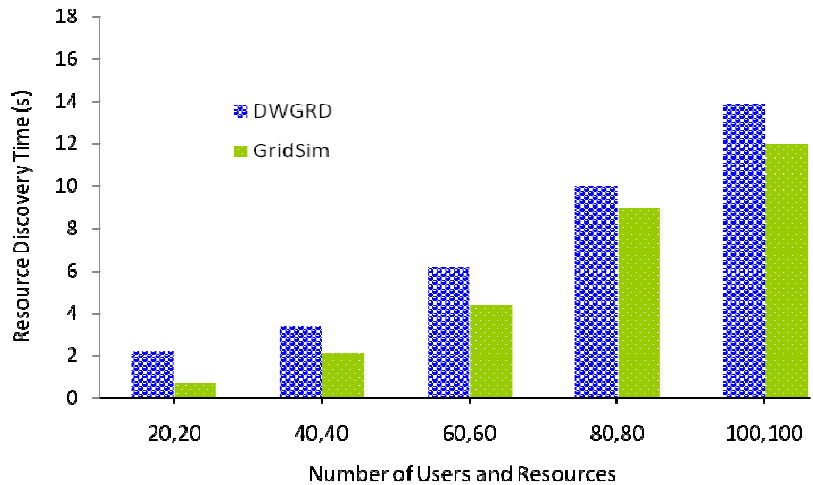

Figure 4.1. Comparison of increase in means of DWGRD and GridSim

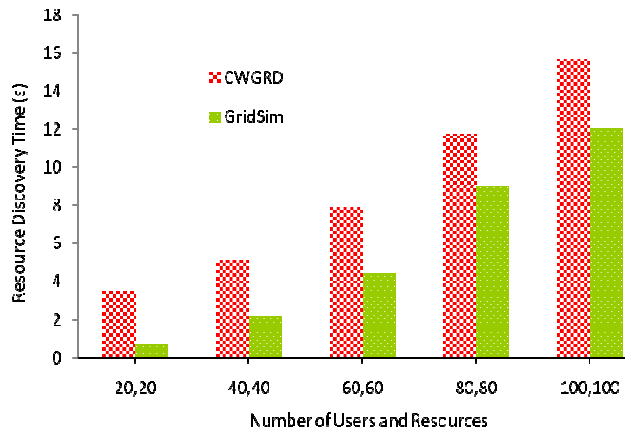

Figure 4.2. Comparison of increase in means of CWGRD and GridSim

As expressed in the previous sections, some of the Grid parameters that change resource discovery time are simulated by GridSim. Other factors are also part of our model and depend on whether resource discovery is done through DWGRD, CWGRD or directly through GridSim. To be able to compare all three models in this section, we first looked at the effect of increasing resources with a fixed numbers of users 20, 60 and 100 with all the same conditions as in the above sections. We also used the same data as given in the above sections for each model. The two-dimensional graphs that demonstrate the results are shown next in Figures 4.3, 4.4 and 4.5 respectively.

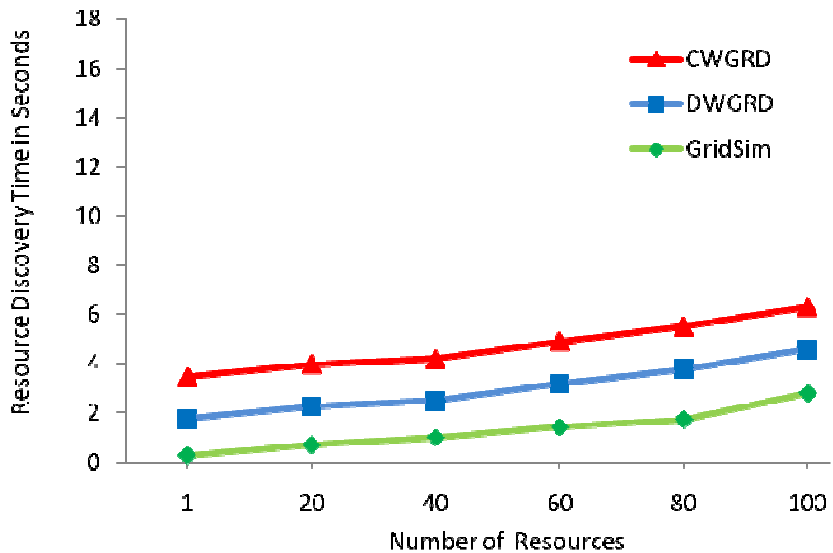

Figure 4.3 Average User Time of GridSim, DWGRD and CWGRD with users=20

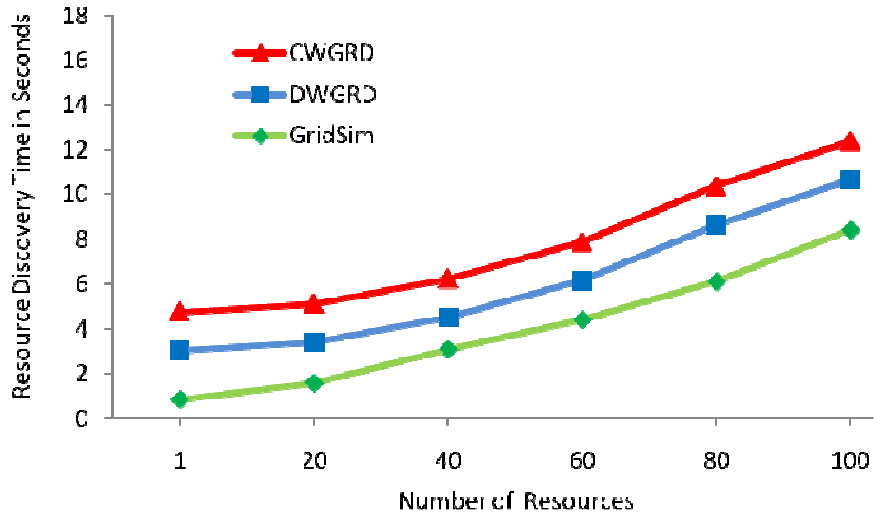

Figure 4.4 Average User Time of GridSim, DWGRD and CWGRD with users=60

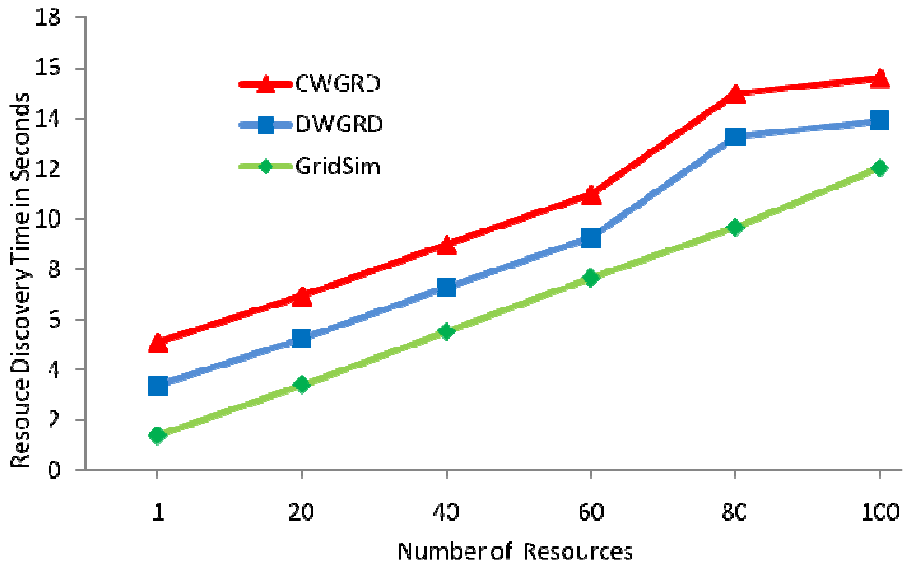

Figure 4.5 Average User Time of GridSim, DWGRD and CWGRD with users=100

The reason for the increase of resource discovery timing is generation of more simulated resources in GridSim (No Architecture), DWGRD and CWGRD. The DWGRD time includes the web service execution time and the time to discover resources through the resource finder web service plus GridSim resource registration time. The CWGRD time includes the time to discover the URL of web service from jUDDI registry that is the time to search the registry and running the application plus the time of DWGRD that includes the time of calling web service plus GridSim resource registering time.

Since we want to test the scalability, increasing the number of users with the fixed number of resources set to 20, 60 and 100 is the other criteria that we considered to observe for analyzing the effects of increasing users on resource discovery time.

A constant increase in resource discovery timing is found as the number of users is increased with resources constant at 20, 60 and 100 in the CWGRD, DWGRD and GridSim models as shown next in Figures 4.6, 4.7 and 4.8.

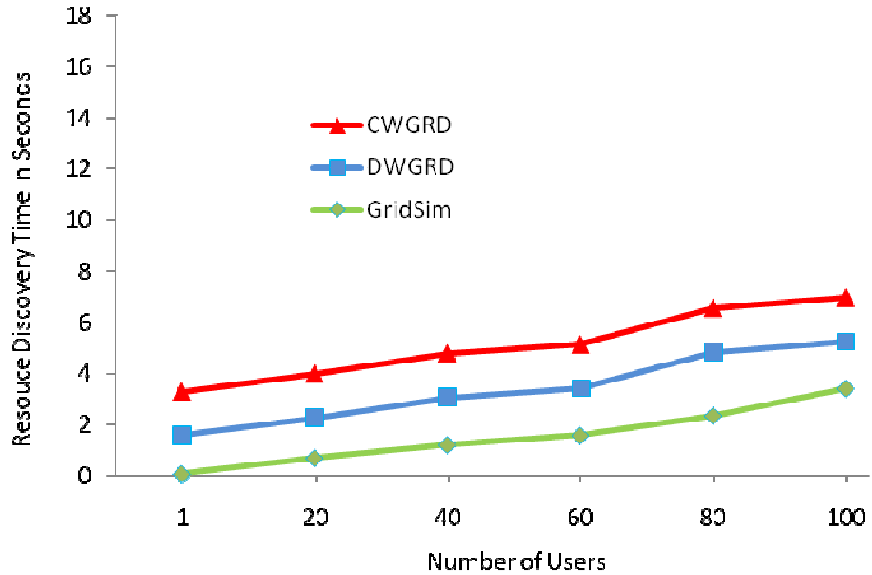

Figure 4.6 Average User Time of GridSim, DWGRD and CWGRD with resources=20

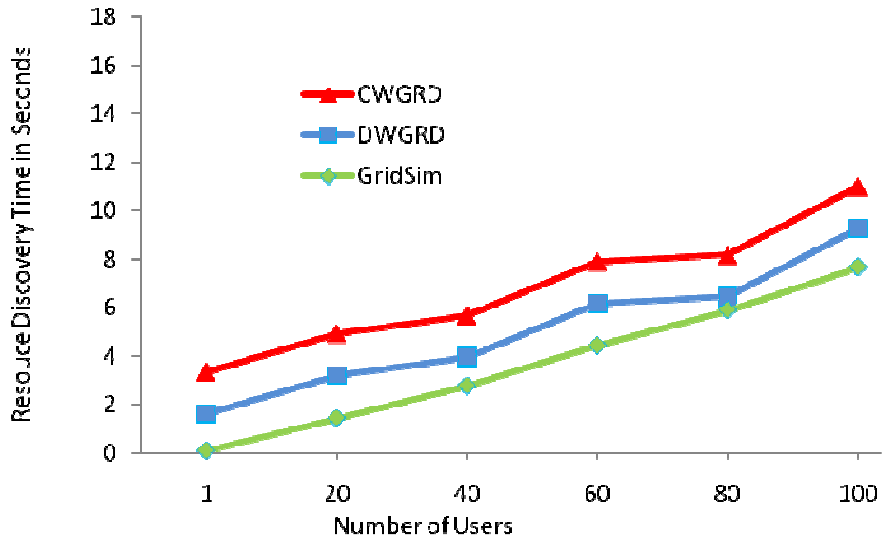

Figure 4.7 Average User Time of GridSim, DWGRD and CWGRD with resources=60

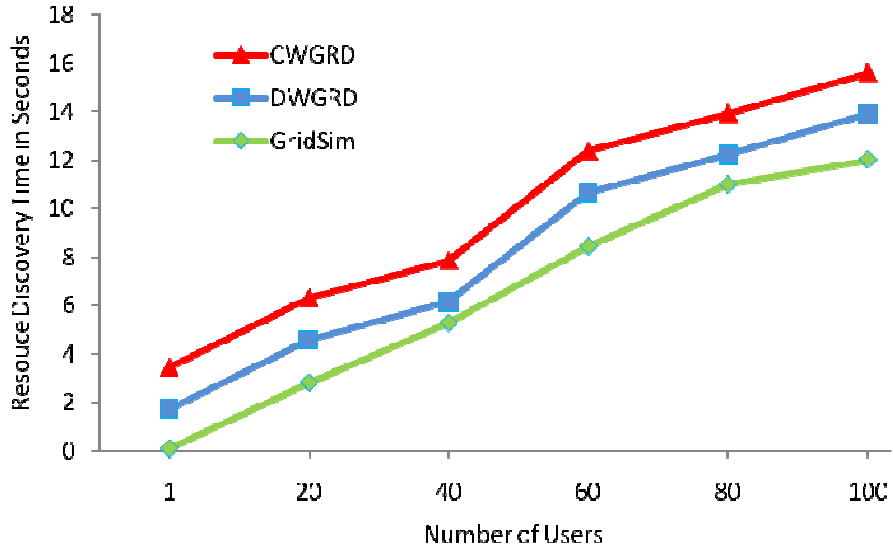

Figure 4.8 Average User Time of GridSim, DWGRD and CWGRD with resource=100

As these graphs show, the average time of resource discovery of an average of 100 users with 100 resources in the CWGRD model is found to be 15.618 seconds. The average time of resource discovery of an average of 100 users with 100 resources in DWGRD model is found to be 13.902 seconds, while that of GridSim is found to be 12.012 seconds. As we can see in figures 4.6 and 4.9, the difference between resource discovery time of GridSim, DWGRD and CWGRD are the same for a fixed 100 users and 100 resources. We also want to see the difference between resource discovery times of three models when increased number of users and resources are equal.

## 5 ANALYSIS

After visual comparison, statistical testing is performed to analyze the difference between these means. Considering the low number of raw data points selected for each sample, we selected t-test for statistical analysis. The results of the statistical testing were confirmed online at [23].

To find the significance of the differences between the means of the resource discovery times of our model (DWGRD and CWGRD) and the resources discovery times achieved by GridSim, we calculated confidence intervals using the t-student distribution and performed t-test for the five points shown in the previous graphs. These points show the linear increase in resource discovery time for 20, 40, 60, 80 and 100 users and resources. Please note that users and resources are equal in number for each point. The t-student distribution is used with the following assumptions:

1. Samples of two populations are independent of each other.

2. Samples are drawn from normal populations.

3. The populations have the same variance.

Since the results of the experiments of CWGRD and DWGRD are independent of GridSim, we did the t-test for unpaired observations and the p-value recorded is shown in Tables 1 and 2. To perform the test, we followed Jain's method [24] and calculated the required statistics for the 10 observations for GridSim, DWGRD and CWGRD. The statistical results shown in Tables 1 and 2 are calculated using the following equations:

Mean
$$\overline{x_{a/b}} = \frac{1}{n_{a/b}} \sum_{i=1}^{n_{a/b}} x_{ia/b}$$
Equation 1

Mean Difference
$$x_a - x_b = \frac{1}{n_a} \sum_{i=1}^{n_a} x_{ia} - \frac{1}{n_b} \sum_{i=1}^{n_b} x_{ib}$$
Equation 2

Standard Deviation
$$S_{a/b} = \left\{ \frac{\left( \sum_{i=1}^{n_{a/b}} x_{ia/b}^2 \right) - n_{a/b} \overline{x_{a/b}}^2}{n_{a/b} - 1} \right\}^{\frac{1}{2}}$$
Equation 3

Standard Deviation of Mean Difference
$$S = \sqrt{\frac{S_a^2}{n_a} + \frac{S_b^2}{n_b}}$$
Equation 4

Degree of Freedom for 2 sample populations
$$(n_a + n_b) - 2$$
Equation 5

Confidence Interval
$$(\overline{x_a} - \overline{x_b}) \pm t_{[1-\frac{\alpha}{2})(n_a+n_b-2)]} S$$
Equation 6

In the equations above, *a* stands for first raw data population and *b* stands for second raw data population.

As shown in Tables 1 and 2, we used these formulas to calculate mean differences, standard deviation of mean differences for finding t-scores, p-values and the confidence intervals. The confidence intervals are calculated based on t-distribution table given in [25].

Table 1. Mean Difference between GridSim and DWGRD Confidence Interval and Unpaired T-test

| Users, Resources | μ Difference | σ of Mean Differences | Confidence Interval 95% | p-Value | Mean Differences |
|---|---|---|---|---|---|
| 20,20 | 1.573 | 0.118 | (1.326, 1.820) | <0.0001 | Different |
| 40,40 | 1.227 | 0.478 | (0.221, 2.232) | 0.0196 | Different |
| 60,60 | 1.744 | 0.980 | (-0.314, 3.802) | 0.0918 | Insignificant |
| 80,80 | 1.025 | 1.864 | (-2.891, 4.941) | 0.5891 | Insignificant |
| 100,100 | 1.890 | 3.721 | (-5.927, 9.707) | 0.6176 | Insignificant |

Table 2. Mean Difference between GridSim and CWGRD Confidence Interval and Unpaired T-test

| Users, Resources | μ Difference | σ of Mean Differences | Confidence Interval 95% | p-Value | Mean Differences |
|---|---|---|---|---|---|
| 20,20 | 3.289 | 0.121 | (3.036, 3.542) | <0.0001 | Different |
| 40,40 | 2.946 | 0.475 | (1.948, 3.944) | <0.0001 | Different |
| 60,60 | 3.462 | 0.977 | (1.410, 5.513) | 0.0023 | Different |
| 80,80 | 2.742 | 1.865 | (-1.176, 6.660) | 0.1587 | Insignificant |
| 100,100 | 3.606 | 3.719 | (-4.208,11.421) | 0.3451 | Insignificant |

Following Jain [24], we calculated the confidence interval for each comparison. According to these tests, if the calculated interval includes 0 or the p-value is greater than 0.05, the difference between two means is not significant.

For calculating the p-value, we used the method outlined by Mendenhall et al. [25]. The confidence intervals and p-values lead to the following conclusions:

1. The difference between the means of resource discovery time of DWGRD and GridSim is significant only when resources and users are less than or equal to 40.

2.  When the number of resources and users are increased to more than 40, the difference between GridSim and DWGRD is not significant.

3.  The difference between the means of resource discovery time of CWGRD and GridSim is significant when the numbers of resources and users are 20, 40 and 60. This is an expected result when considering the greater value of the mean of CWGRD resource discovery time compared to the mean of DWGRD resource discovery time. We expect to see a greater difference between CWGRD to GridSim than DWGRD to GridSim.

Finally, the results of the t-test support the scalability of our model. In the simulation, the scalability can be tested when we load the grid with more resources and users. Therefore, we expect that the cost (i.e., the resource discovery time) of adding our model compared to the resource discovery time of grid to be marginal when accessing the resource grids through the Internet.

# 6 CONCLUSIONS AND FUTURE WORK

In this work, a web service-based layered model that utilizes a distributed search for resource discovery in grid computing is presented. This model provides a scalable solution for information administrative requirements when the grid system expands over the Internet. Distributed search in this model is based on sending DNS queries to the repositories that are distributed in the Internet. Two architectures required for regional grid resource discovery in the model are implemented: Direct and Centralized Web-Based Grid Resource Discovery.

These two architectures act as the best-case scenarios of the model because they simulate a successful regional search for the URL of resource finder service which is used in the model. The DWGRD architecture implemented in this work is obviously the best-case scenario of our model. In DWGRD, the client knows the resource finder's URL from the beginning.

CWGRD implementation is the second implemented architecture that is based on contacting central UDDI database registry located in the same region as the client. We consider this case the second-best case scenario because we assume the URL of regional UDDI is known by the client and UDDI successfully returns the URL of the required resource finder web service to the client.

The slightly higher average time for resource discovery observed in the DWGRD model compared to the GridSim is because of the additional time for using web services to send the simulation parameters and initiating the proper classes to start GridSim.

The average resource discovery times of CWGRD are higher in comparison to GridSim and DWGRD. Because CWGRD searches the URL of web service from a central regional database, it then behaves similar to the DWGRD model. The simulation for finding the total grid resource discovery time is conducted in the same environment for both DWGRD and CWGRD models.

Whenever the URL of the resource finder cannot be found locally, it means there is no resource finder that satisfies client's requirements within the current region. In this case, our model must perform a distributed search. In the distributed search, our model uses caching in UDDI repositories, which is why searching time in our model after the first time should be equal to the best-case scenarios (i.e., regional search) in the rest of the cases.

We found that when increasing the load of the grid, the resource discovery time (i.e., cost) of our model becomes negligible. With a load of 100 users and 100 resources, there is no significant difference between the resource discovery time of our model and no model (GridSim).

Based on this result, we conclude that if our model were implemented on the Internet with thousands of grid resources and users, the cost of our model is negligible, which means the model would be scalable.

The limitation of this work is that both the DWGRD and CWGRD architectures are implemented with one client and one server on the same machine as the grid simulator. This work in future can be extended to implement the model with multiple clients and multiple server nodes examining a variety of scenarios of finding the resources distributed in a real grid computing environment.

Also, research on the implementation and simulation of other parts of the grid in the lower layers to enhance our model is the future direction of this work.